\documentstyle[prl,aps,twocolumn,floats,tighten,fleqn,epsfig]{revtex}

\begin{document}

\draft

\twocolumn[
\hsize\textwidth\columnwidth\hsize\csname@twocolumnfalse\endcsname

\title{Possible Indications of a Clumpy Dark Matter Halo}

\author{Lars Bergstr\"om}  
\address{Department of Physics, Stockholm University, Box 6730,
SE-113~85~Stockholm, Sweden}

\author{Joakim Edsj\"o}
\address{Center for Particle Astrophysics, University of California,\\
301 Le Conte Hall, Berkeley, CA 94720-7304, U.S.A}

\author{Piero Ullio}
\address{Department of Physics, Stockholm University, Box 6730,
SE-113~85~Stockholm, Sweden}

\maketitle

\begin{abstract}

We investigate if the gamma ray halo, for which recent evidence has
been found in EGRET data, can be explained by neutralino annihilations
in a clumpy halo. We find that the measured excess gamma ray flux can
be explained through a moderate amount of clumping in the halo.
Moreover, the required amount of clumping implies also a measureable
excess of antiprotons at low energies, for which there is support from
recent measurements by the BESS collaboration.  The predicted
antiproton fluxes resulting from neutralino annihilations in a clumpy
halo are high enough to give an excess over cosmic-ray produced
antiprotons also at moderately high energies (above a few GeV). This
prediction, as well as that of one or two sharp gamma lines coming
from annihilations into $\gamma\gamma$ or $Z\gamma$ can be tested in
upcoming space-borne experiments like AMS and GLAST. 
 
\end{abstract}


]

\narrowtext



Some models for structure formation in the universe predict that
cold dark matter clumps may have formed at various stages of the
evolution of structure. Subsequent hierarchical merging of small
clumps into larger ones would evenually give rise to halos of
galaxies. However, if small and dense enough, some of these dark
matter clumps may have survived tidal interactions and could exist
still today in the halos of galaxies including the Milky Way.  A
clumpy halo composed of supersymmetric dark matter particles
(neutralinos) would reveal itself in conspicuous ways due to the
increased annihilation rate of neutralinos into gamma rays, positrons,
neutrinos and antiprotons.

Several authors \cite{dixon,hunter,bertsch,sreekumar} have pointed out
that there is a discrepancy between the measured diffuse $\gamma$ ray
background in the Milky Way and the predictions of detailed emission
models. A strong excess of photons with energy above 500 MeV has recently been
detected towards the galactic center \cite{M-H} by the Energetic Gamma Ray 
Experiment Telescope (EGRET) on board the Compton Gamma Ray
Observatory. Rather significant deviations from emission models in the
same energy range seem to be present on a large scale as well. In a
very recent analysis, Dixon et al. \cite{dixon} (hereafter Paper I)
claim that in EGRET data there is a strong statistical evidence for a
$\gamma$ ray halo surrounding the galaxy.  

Assuming that the effect is real, some possible explanations have been
addressed, such as an origin from unresolved point sources
(Geminga-like pulsars are feasible candidates), an underestimate of
the inverse Compton emission at high latitudes and $\gamma$ rays
associated with baryonic dark matter or WIMP annihilations. We will
focus on the latter and investigate the intriguing possibility that
the $\gamma$ ray halo might result from pair annihilations of dark
matter neutralinos $\chi$, the lightest supersymmetric particle in the
Minimal Supersymmetric Standard Model (MSSM) and one of the leading
dark matter candidates. We will analyse the compatibility of the
signal with this hypothesis and compare with the information which
could be extracted from other neutralino detection
methods, in particular low-energy cosmic antiprotons. Recent results 
from the Balloon-borne Experiment with Solenoidal magnet Spectrometer
(BESS) \cite{bess} indeed appear to show some excess of antiprotons 
at low energy, 
a characteristic signal of neutralino annihilations.


We work in the framework of the Minimal Supersymmetric Standard Model
(MSSM) as defined in Refs.~\cite{haberkane,jkg}. More details on our
notation can be found in Ref.~\cite{coann}.  With some simplifying
assumptions we are left with 7 parameters which we allow to be varied
within generous bounds.  The ranges for the parameters are given in
Table~\ref{tab:scans}.  
For each generated model, we check if it is
excluded by current accelerator constraints and if it is
cosmologically interesting, by which we mean models where $0.025 <
\Omega_{\chi}h^{2} <1$, i.e.\ where the neutralinos can make up most
of the dark matter in our galaxy without overclosing the Universe. ($\Omega$
is the energy density in units of the critical density and the present
Hubble parameter is $100h$ km s$^{-1}$ Mpc$^{-1}$.)


\begin{table}[!t]
  \centering 
  \begin{tabular}{lrrrrrrr} 
  Parameter & $\mu$ & $M_{2}$ &
  $\tan \beta$ & $m_{A}$ & $m_{0}$ & $A_{b}/m_{0}$ & $A_{t}/m_{0}$ \\
  Unit & GeV & GeV & 1 & GeV & GeV & 1 & 1 \\ \hline 
  Min & -50000 & -50000 & 1.0  & 0     & 100   & -3 & -3 \\
  Max & 50000  & 50000  & 60.0 & 10000 & 30000 &  3 &  3 \\ 
  \end{tabular} 
\caption{The ranges of parameter values used in our scans of the MSSM
  parameter space. Note that several special scans aimed at
  interesting regions of the parameter space has been performed.
  In total we have generated about 55000 models that are not excluded by
  accelerator searches.}  \label{tab:scans}
\end{table}



To get the normalization of the $\gamma$ ray flux from neutralino 
annihilations, we need to specify the distribution of dark matter 
in the galaxy. We first assume that the average dark matter density profile 
is spherically symmetric and can be described by
\begin{equation}
  \rho(r) = \rho_0\; \left(\frac{R_0}{r}\right)^{\gamma} 
  \,\left[\frac{1+(R_0/a)^\alpha}{1+(r/a)^\alpha}\right] 
  ^{(\beta-\gamma)/\alpha}
  \label{eq:halo}
\end{equation}
where $\rho_0$ is the local halo density, $R_0$ our galactocentric 
distance and $a$ some length scale.  The functional form of $\rho$ for 
the Milky Way cannot be fully determined from observational 
data~\cite{binney} and several choices of the parameters are 
acceptable.  A hint on this choice may come from N-body simulations of 
hierachical clustering in cold dark matter cosmologies which favour 
profiles that are singular towards the galactic centre.  We will 
consider the Navarro et al.\ profile~\cite{navarro} which is cuspy 
and has $\gamma = 1$ (it scales as $1/r^{3}$ at large distances) and, for 
comparison, the modified isothermal distribution, 
$(\alpha,\beta,\gamma)=(2,2,0)$, extensively used in dark matter 
detection computations.  We have checked that the Kravtsov et al.\ 
profile~\cite{kravtsov} which is mildly singular with $\gamma \sim 
0.2$, also gives acceptable fits.

To make the analysis compatible with the existence of
clumps in the halo, we consider the possibility that a fraction $f$ 
of the total dark matter, rather than being smoothly distributed in 
the halo, is concentrated in clumps. 
Simulations of structure formation in the 
early Universe do not yet have the dynamical range to give 
predictions for the size and density distribution of small mass 
clumps (we focus here on clumps of less than around $10^6$ solar 
masses which avoid the problem of unacceptably heating the 
disk~ \cite{silkstebbins}). 
The formation of clumps on all scales is however a 
generic feature of cold dark matter models which have power on all 
length scales. If self-similarity is a guide, galaxy halos may form 
hierarchically in a similar way to that of cluster halos (see e.g.\ 
\cite{bepi}). 
The main effects of a clumpy halo can be sketched by simply
assuming that clumps of given mass have an enhanced 
and roughly constant dark matter density $\rho_{cl}$. 
We introduce the parameter $\delta$ as the contrast between 
the dark matter density in clumps and the 
local halo density which is about 0.3 GeV cm$^{-3}$.
We further assume that the average clump mass distribution follows 
the same distribution as the smooth component, Eq.~(\ref{eq:halo}).


Pair annihilations of neutralinos in the dark matter halo can produce
photons which are either monochromatic or with a continuum energy
spectrum. The monochromatic $\gamma$s arise from the loop-induced
S-wave annihilation into the $\gamma\gamma$ or $Z \gamma$ final states;
the phenomenology of these processes has been studied recently in
Ref.~\cite{lpj}. The continuum contribution is on the other hand
mainly obtained from pions produced in jets; these photons are
expected to be much more numerous but lower in energy than the
monochromatic ones \cite{continuum}. To model the fragmentation 
process and extract
information on the number and energy spectrum of the photons produced
we have used the Lund Monte Carlo {\sc Pythia} 6.115~\cite{pythia}.

Consider a detector with an angular 
acceptance $\Delta\Omega$ pointing in a direction which forms an 
angle $\psi$ with respect to the galactic centre.
The integrated $\gamma$ ray flux above an energy threshold $E_{\rm th}$
is given by
\begin{eqnarray}
  \lefteqn{\Phi_\gamma (E_{\rm th},\,\Delta \Omega,\, \psi) 
  \simeq 1.87 \cdot 10^{-8} 
  \, {\cal S} (E_{\rm th}) \, \cdot} \nonumber \\
  & & \langle\,J\left(\psi \right)\,\rangle\,(\Delta\Omega)
  \;\;\rm{cm}^{-2}\;\rm{s}^{-1}\;\rm{sr}^{-1}\;. 
  \label{eq:flux}
\end{eqnarray}
In this formula we have defined a particle physics dependent term
\begin{eqnarray}
  {\cal S} (E_{\rm th}) &=& 
  \left( \frac{10\,\rm{GeV}}{M_\chi}\right)^2
  \int_{E_{\rm th}}^{M_\chi}\;dE\, \cdot \nonumber \\
  & & \cdot\sum_F \left( \frac{v\sigma_F}{10^{-26}\ {\rm cm}^3 
   {\rm s}^{-1}}\right)\,\frac{dN_\gamma^{\,F}}{dE}
\end{eqnarray}
where $M_\chi$ is the neutralino mass, $F$ are the allowed final 
states and for each of these $v\sigma_F$ is the annihilation rate, 
while $dN_\gamma^{\,F}/dE$ is the differential distribution 
of produced photons. The dependence of the flux on the dark matter 
distribution is contained in the factor 
$\langle\,J\left(\psi \right)\,\rangle\,(\Delta\Omega)$. 
In a scenario with many unresolved clumps, it 
is possible to show that the contribution from the clumps is given 
by~\cite{clumpy}
\begin{eqnarray}
  \lefteqn{\langle\,J\left(\psi \right)\,\rangle_{\rm cl}\,
  (\Delta\Omega) 
  = \frac{1} {8.5\, \rm{kpc}}\,\frac{1} {\Delta\Omega}
  \; f \delta} \nonumber \\
  & & \int_{\Delta\Omega}\,d\Omega'\;\int_{\rm line\;of\;sight}\,dl\;
  \left(\frac{\rho(l,\,\psi')}
  {0.3\,{\rm GeV}/{\rm cm}^3}\right)
\end{eqnarray}
in contrast to the smooth case \cite{lpj} where the dependence is quadratic
in the density $\rho$.
The relative strength of the smooth and the clumped components 
is determined both by 
the halo profile and by the parameter $f\delta$, the product of the 
halo fraction in clumps and their overdensity. 


\begin{figure}
  \centerline{\epsfig{file=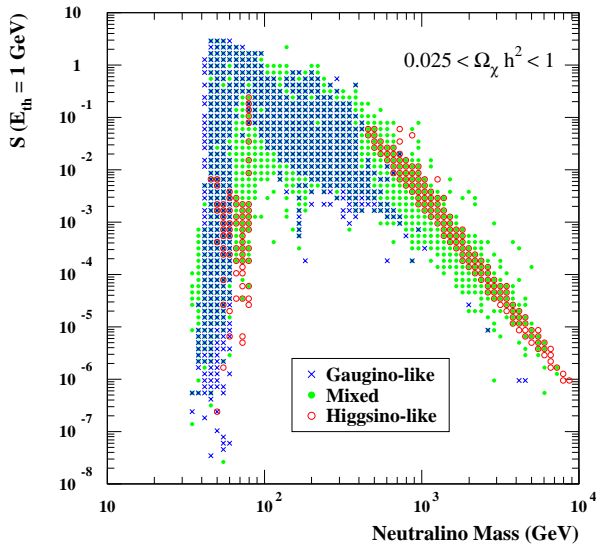,width=0.49\textwidth}}
  \caption{The value of ${\cal S}$ versus the neutralino mass.}
  \label{fig:svsm}
\end{figure}

In Fig.~\ref{fig:svsm} we show the values of ${\cal S}(1\ {\rm GeV})$ 
versus the neutralino mass for our set of supersymmetric models. 
Our results show a very large dispersion for the possible values of 
${\cal S}$, about seven orders of magnitude; the highest ${\cal S}$ 
are for low neutralino masses, $M_\chi \sim 40-60$ GeV, while for 
heavier neutralinos there are both the $1/M_\chi^2$ suppression and 
lower photon production rates. The highest values of the $\gamma$ 
ray flux are given by gaugino-like neutralinos and mixed neutralinos, 
whereas ${\cal S}$ is generally much lower for higgsino-like 
neutralinos, at least in the low mass range (the opposite trend was 
noticed for the monochromatic photon flux, see Ref.~\cite{lpj}). 
For the highest 
possible values of the flux  ${\cal S}$ 
scales as $1/(\Omega_\chi h^2)$ which reflects the fact that to 
a first approximation $\Omega_\chi h^2 \propto 1 / (v\sigma)$. 
Therefore, the maximal ${\cal S}$ depends crucially on the minimal 
$\Omega_\chi h^2$ (which we take to be 0.025) 
that we judge to be cosmologically interesting. 
Current estimates of $\Omega$ and $h$ indicate (with large errors)
$\Omega h^2\sim 0.1$, of which baryons may contribute around 
$0.02-0.03$ \cite{schramm}.

We focus now on the result reported in Paper I. The 
value of the residual flux at high latitudes shown in Fig.~3, Paper I 
is about $10^{-6}$ ph cm$^{-2}$ s$^{-1}$ sr$^{-1}$. Even picking the 
MSSM model in our set which gives the highest 
${\cal S}(1\ {\rm GeV})$, 
it is not possible to reproduce the qualitative features in that 
Figure in a smooth halo scenario with any of the three halo models 
considered and any of the allowed values for length scale $a$ and 
local halo density $\rho_0$. On the other hand a many unresolved 
clump scenario may be compatible with the results of Paper I. We will 
show this through two examples. 

\begin{figure}
  \centerline{\epsfig{file=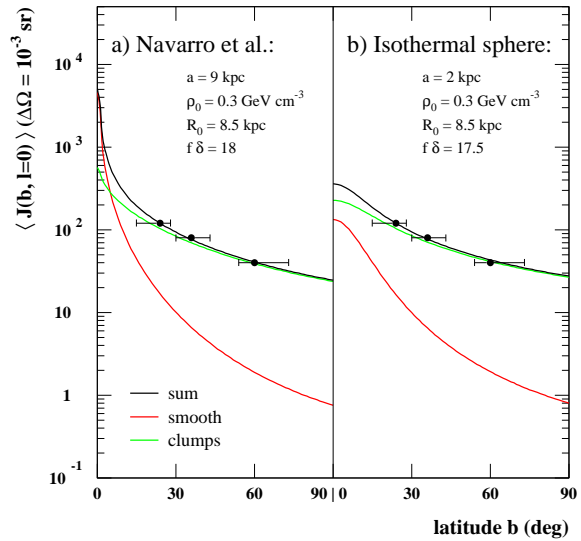,width=0.49\textwidth}}
  \caption{The value of $J(\psi)$ for a) the Navarro et al.\ profile 
   and b) the isothermal sphere.
  The clumpyness and halo model parameters to fit the EGRET data, shown
  as filled circles,
  are given in the figure.}
  \label{fig:jpsi}
\end{figure}

Let us first consider an example in the low mass range, 
a MSSM model which has $M_\chi = 76$
GeV, $\Omega_\chi h^2 = 0.03$, $Z_g = 0.18$ and which gives ${\cal
S}(1\ {\rm GeV}) = 1.025$. 
We show in Fig.~\ref{fig:jpsi} for this model the values of 
$\langle\,J\left(\psi \right)\,\rangle$ needed to fit the
results in Paper I. We are considering a roughly spherical 
$\gamma$ ray halo and deriving the angular distribution from 
Fig.~3, Paper I at zero longitude.
As can be seen, the present EGRET data can be quite well reproduced 
in our approach. Although we are not in the position of 
discriminating among different halo profiles, one may notice that 
in the case of the Navarro et al.\ profile, a sharp enhancement of 
the $\gamma$ ray flux is predicted towards the galactic centre, due 
to the singularity of the smooth density profile. This is a feature 
which may be searched for in the EGRET data (some indications of 
results going in that direction were given in Ref.~\cite{M-H}).

We find that the dark matter signal, which causes an excess
of photons mainly in the energy range between a few GeV and
$M_\chi$, may explain
the discrepancy between the 2.75 power law fall off which is expected
according to models of diffuse galactic background and the behaviour
that has been mapped by EGRET up to 20 GeV. Future 
experiments~\cite{ams,glast} will be
able to measure the diffuse background at much higher energies and
eventually detect if there is a break in the energy spectrum at about
the neutralino mass. 
 
For the MSSM model we have chosen, the gamma lines from 
annihilation into $\gamma\gamma$ and $Z\gamma$ are well above the
background, and their detection might be possible, especially if there
is an enhancement of the dark matter density towards the galactic
centre. The detection of lines, which have no plausible astrophysical
background, seems to be the natural way to show conclusively whether
the $\gamma$ ray halo originates from dark matter neutralino
annihilations. On the other hand, the correlation between continuum
and line signal strengths is very weak and the absence of the lines 
would by no mean imply that there cannot be a continuum signal.

Given the value of $f\delta$ needed,
 we have to worry about other dark matter searches
and make sure that these models are not excluded already. It turns out
that the signals are below present detection limits, except for the
antiproton signal which we now discuss.

We have computed the antiproton signal in much the same way as the 
continuous gammas by using the Lund Monte Carlo {\sc Pythia} 6.115.  
We have applied for the propagation the leaky box approximation with 
the energy dependent escape time given in Ref.\ \cite{salati-pbar} and 
used the solar modulation model of Ref.\ \cite{solarmod}.

For the example considered above the antiproton flux at 0.4 GeV in a 
clumpy halo with $f \delta=18$ is $\phi_{\bar{p}} = 1.6 \times 
10^{-5}$ cm$^{-2}$ s$^{-1}$ sr$^{-1}$ GeV$^{-1}$ which should be 
compared with the flux measured by BESS\cite{bess} $\phi_{\bar{p}} = 
1.4^{+.9}_{-.6} \times 10^{-6}$ cm$^{-2}$ s$^{-1}$ sr$^{-1}$ 
GeV$^{-1}$ at the same energy.  It may be tempting to conclude that 
this model is already excluded since it gives a too high antiproton 
flux, but one has to keep in mind the big uncertainties involved, 
mainly in the antiproton propagation.  In particular, it is not clear 
how large a fraction of antiprotons generated in the halo (i.e.\ 
outside the galactic disk) can penetrate the wind of cosmic rays 
leaving the disk \cite{ptuskin}.  It is interesting that the 
antiproton flux is within an order of magnitude of the reported BESS 
flux, which shows some of the characteristics (pile-up at low energy) 
expected for neutralino-induced antiprotons \cite{bess}.

Since we have found in our calculations that the antiproton flux 
strongly correlates with the continuous gamma flux for a given 
supersymmetric model, it seems impossible to reduce the antiproton 
flux maintaining high continuous gamma flux at sub-100 GeV neutralino 
mass.  If the overproduction of antiprotons seems uncomfortably high, 
it is however possible to resolve this by going to higher neutralino 
masses.

We thus choose as our second example an MSSM model which has a large 
mass, $M_\chi = 503$ GeV, $\Omega_\chi h^2 = 0.03$, $Z_g = 0.04$ and 
which gives ${\cal S}(1\ {\rm GeV}) = 0.05$ ph.  The necessary 
rescaling for this model is $f \delta=427$ for which the antiproton 
flux at 0.4 GeV is $\phi_{\bar{p}} = 1.7 \times 10^{-6}$ cm$^{-2}$ 
s$^{-1}$ sr$^{-1}$ GeV$^{-1}$, i.e.\ within the $1\sigma$ error bars 
of the BESS measurement.  For these higher mass models, there may be a 
problem in explaining a high gamma ray flux in the lower energy 
interval 0.3--1.0 GeV (this interval is however subject to larger 
uncertainties).  An intermediate mass model in which the antiproton 
flux limit by BESS is mildly violated might of course be considered.  
More data is obviously needed to make firmer statements.

We remark that for this particular model, the spin-independent cross
section on nucleons is $0.15\cdot 10^{-4}$ pb, very close to the limit
given by the most sensitive $NaI$ experiment \cite{dama}. 
For other models giving similar gamma and antiproton rates, however, 
the direct detection rates are much lower.

\begin{figure}
  \centerline{\epsfig{file=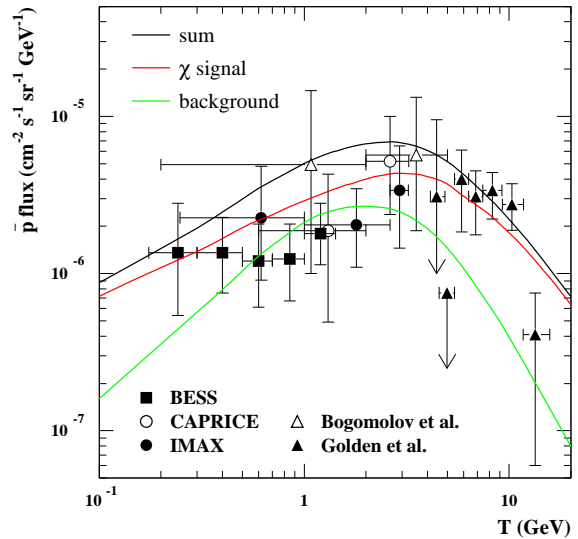,width=0.49\textwidth}}
  \caption{The energy spectrum of antiprotons for the second 
  model example (see text for details).}
  \label{fig:pbar}
\end{figure}

It is interesting to note that for the high-mass neutralino,
also a high-energy excess of 
antiprotons is potentially measurable. This is illustrated in 
Fig.~\ref{fig:pbar}, where a compilation of present 
data (\cite{bess} and references therein) is shown 
together with the predictions of cosmic-ray induced 
background (the mid-range of the predictions of \cite{pbarback}), and 
the flux in our second example. As can be seen, present data are not 
yet conclusive. However, an interesting 
feature of the high-mass neutralino result is that the maximum of the 
antiproton flux is shifted towards higher energies by 1--2 GeV
compared with the low mass case. 
Also, the fall-off with energy above the peak is 
considerably slower than for the background. Indeed, our second model 
fits quite nicely this higher-energy part of the present data. It 
should be noted that at these
 energies, the effects of galactic and solar wind modulation is 
less severe than at sub-GeV energies, making the predictions more 
trustworthy. These features should definitely be 
investigated in the upcoming antiproton 
measurements~\cite{pamela,ams}. 

Although the interpretation of the measured excess in cosmic gamma 
rays and antiprotons in terms of neutralino annihilation contains 
elements of speculation at the present time, it is reassuring that 
upcoming 
experiments will be in a position to more firmly confirm or rule out 
this hypothesis. For instance, the proposed 
Gamma-ray Large Area Space Telescope (GLAST)~\cite{glast} 
will have spectral and angular resolution enough to 
search for gamma ray lines in the direction of the center of the 
galaxy. Also, if the explanation lies in a clumpy halo, a 
large-exposure experiment like GLAST eventually may resolve
individual large clumps as bright gamma-ray spots on the sky. 
The antiproton spectrum will soon be measured with higher accuracy 
as well in the Alpha Magnetic Spectrometer (AMS)~\cite{ams} 
and PAMELA~\cite{pamela} experiments. 
Also, upcoming direct detection experiments may in favourable cases be 
sensitive to these dark matter candidates. Finally, improved $N$-body
simulations of structure formation in cold dark matter models
may give a better assessment of the credibility of clumpy halo models.

\smallskip
{\bf Acknowledgments}
\smallskip

We wish to thank Paolo Gondolo, Ted Baltz and Markku 
J{\"a\"a}skel{\"a}inen for useful discussions. 
L.B. was supported by the
Swedish Natural Science Research Council (NFR). This work was
supported with computing resources by the Swedish Council for High
Performance Computing (HPDR) and Parallelldatorcentrum (PDC), Royal
Institute of Technology.


\end{document}